# ENERGY INNOVATIONS SMALL GRANT TRANSPORTATION PROGRAM

## FINAL REPORT

### Eco-Routing Navigation System for Electric Vehicles

**EISG AWARDEE**

Center for Environmental Research and Technology,
University of California, Riverside,
1084 Columbia Ave., Riverside, CA 92507
Phone: (951) 781-5630
Email:gywu@cert.ucr.edu

**AUTHOR(S)**

Guoyuan Wu, Principal Investigator
Matthew Barth
Kanok Boriboonsomsin

Grant #: 11-01TE
Grant Funding: $94,993
Term: October 2012- March, 2014
Subject Area: Vehicle Technologies

## Legal Notice

This report was prepared as a result of work sponsored by the California Energy Commission (Commission). It does not necessarily represent the views of the Commission, its employees, or the State of California. The Commission, the State of California, its employees, contractors, and subcontractors make no warranty, express or implied, and assume no legal liability for the information in this report; nor does any party represent that the use of this information will not infringe upon privately owned rights. This report has not been approved or disapproved by the Commission nor has the Commission passed upon the accuracy or adequacy of the information in this report.

Inquires related to this final report should be directed to the Awardee (see contact information on cover page) or the EISG Program Administrator at (619) 594-1049 or email eisgp@energy.state.ca.us.

# Acknowledgement Page

The authors would like to thank the California Energy Commission for the sponsor of this project and the Project Manager for her kind help and support. The authors are also grateful to Mike Todd and Alex Vu from the Center for Environmental Research and Technology (CERT) for acquisition of the testing vehicle and database setup. The authors are very thankful for the help and support from Jiwoong Kwak, Shawn Miata, Nicholas Jarak and Junhyuk Kang with CERT on field data collection.

# Table of Contents

# List of Figures

# List of Tables

# Abstract

The goal of this project is to determine the feasibility of calculating a travel route for an electric vehicle (EV) that will require the least amount of energy for the trip, and thus extending the range of the EV. To achieve this goal, the research team set up several objectives, including 1) to demonstrate the accuracy of measurements from the data acquisition system; 2) to demonstrate the validity of the proposed algorithm in estimating energy consumption of the test EV (2013 NISSAN LEAF) in real-time; and 3) to demonstrate the benefit from the prototype system in terms energy savings for the test EV. The results from the project show that the proposed system can fulfill the aforementioned objectives and provide up to 51% of energy reduction, compared with the conventional navigation strategies (i.e., shortest travel distance and least travel duration).

# Executive Summary

Introduction

As the concern over the environmental impact of the conventional petroleum-based vehicles increases, the development of electric vehicles (EVs) receives more and more attention. However, one of the major obstacles to the mass adoption of EVs is "range anxiety"—the fear that an all-electric vehicle will not make it to desired destination before running out of power. Innovations in eco-friendly intelligent transportation system (ECO-ITS) technologies, including the proposed eco-routing navigation, can help address this concern to some extent by enhancing transportation system efficiency, and in effect, reducing energy consumption.

Project Objectives

There are four major objectives within the original scope of works:
1. Demonstrate that the calibrated simulation tool is able to calculate energy consumption of the test EV within an average error of ± 10%.
2. Demonstrate that the distance-based energy consumption model is able to estimate energy consumption of the test EV in real-time within an average error of ± 10%.
3. Perform field tests on an EV with prototype system (i.e., the eco-routing algorithm) versus without prototype system on both freeway and arterial roadways with various gradients for at least 50 trips.
4. Demonstrate that the prototype system saves the energy consumption of the test EV by an average of at least 10%.

Due to the availability of advanced data acquisition system and unexpected delays in obtaining a test electric vehicle, the research team modified the first objective as follows:
1. Demonstrate that the data acquisition system is able to provide measurements (e.g., second-by-second speed) within an average error of ± 10%.

Project Outcomes

Here are the outcomes corresponding to each aforementioned objective:
- For objective 1: As aforementioned, the research team has modified the original objective to a new one. Regarding the new objective, the average errors for measurements from the GPS data logger are less than 1.7% across all study sites.
- For objective 2: By analyzing the data from the field testing, the research team identified that the average relative error is around 0.6% in terms of trip-based energy consumption.
- For objective 3: The research team performed field tests on the EV (2013 NISSAN LEAF) on both freeway and arterial roadways with gradients for more than 100 trips. The simulation runs for comparison include over 4000 trips to evaluate the benefits of the prototype system.

- For objective 4: The extensive tests over different study sites show that the proposed eco-routing navigation algorithm can reduce the trip-based energy consumption by 5 – 25% and 25 – 51%, respectively, compared to the shortest travel distance algorithm and least travel duration one.

## Conclusions

1. The data acquisition systems implemented in this project, including the CONSULT III plus Kit and GPS data logger, are satisfactory.
2. The prototype system has exhibited good performance in estimating the trip-level energy consumption by validated against the measurements directly from the test EV. In addition, energy consumption for an EV may vary greatly with the roadway grade due to the regenerative braking.
3. The number of trips identified in the scope of work is enough and the data set is reliable for algorithm development and validation.
4. Extensive evaluation tests have shown that the proposed prototype system provides significant savings in energy consumption, compared to conventional navigation systems, which has indicated significant potential for commercialization.

## Recommendations

Based on the results from this project, here are a couple of recommendations for the next step:
1. More extensive field data collection is necessary to quantify the impacts of a variety of influential factors, such as the vehicle model, driver behavior, and roadway grade profile.
2. A more generalized system analysis framework is preferable for a larger scale testing. The research team can perform a more extensive evaluation by gridding a larger network and defining the origin-destination pair based on grids.
3. It is also favorable to analyze the impacts of time of the day or day of the week on different route choice if the information is available.
4. For commercialization, how to select a more rational navigation objective function should receive more attention in order to achieve the balance between energy consumption and travel time or travel distance and fulfill the customers' needs.

## Public Benefits to California

If taking the NISSAN LEAF as an example, implementation of the proposed eco-routing navigation system for EVs can annually save approximately 20,736,000 kWh in electricity, or $4.15 million (assuming $0.2 per kWh) in electricity costs in California. The energy and cost savings will be even greater, if EVs from other manufacturers also adopt the proposed eco-routing navigation technology.

# Introduction

In the last few decades, increased concern over the environmental impact of the petroleum-based vehicles, along with the peak oil, has led to renewed interest in the development of electric vehicles (EVs) [Eberle & Helmolt, 2010]. A key feature which distinguishes EVs from fossil fuel-powered vehicles is that the electricity they consume can be generated from a wide range of sources, including fossil fuels, nuclear power, and renewable sources such as tidal power, solar power, wind power, or any combination of those. Compared with fossil fuel-powered vehicles, EVs can be powered by electricity generated from more efficient sources and emit no tailpipe carbon dioxide ($CO_2$) or pollutants such as oxides of nitrogen ($NO_x$), non-methane hydrocarbons (NMHC), carbon monoxide (CO), and particulate matter (PM) at the point of use [European Commission, 2011].

In addition to the development of vehicle technologies, several innovations in eco-friendly intelligent transportation system (ECO-ITS) technologies have emerged in the last several years, which can enhance transportation system efficiency, and in effect, reduce fuel consumption. These ECO-ITS technologies are aimed at improving vehicle operation by smoothing the stop-and-go driving, navigating vehicles to avoid congestion and steep road grade, and designing new energy management strategies for electric drive systems that optimize based on knowledge of the trip [Wu et al., 2010]. It has been shown that each ECO-ITS technology can potentially reduce fuel consumption and greenhouse gas (GHG) emissions on the order of 5–15% for conventional fossil fuel-powered vehicles [Barth & Boriboonsomsin, 2008]. One example of such technology is an eco-routing navigation system, which can find a travel route requiring the least amount of fuel or energy instead of the shortest-distance or shortest-duration route [Boriboonsomsin et al., 2010].

In the last decade, there has been a proliferation of GPS-guided navigation systems that assist drivers on which routes to take to their intended destinations. For most off-the-shelf navigation systems, they attempt to minimize distance traveled. It is important to point out the shortest-distance route, however, will note minimize energy consumption and emissions in many real-world cases. For example, the shortest-distance route may include heavily congested roadways sections, resulting in longer time spent and higher energy consumed. Newer generations of navigation systems take into account speed limits or even real-time traffic conditions and provide drivers the shortest-duration routes, which may have vehicles travel longer distances, albeit on less congested roadways. Nevertherless, traveling at high speeds for longer distances may result in higher energy consumption (and emissions) compared to a more direct route at lower speeds [Barth et al., 2007; Ahn and Rakha, 2008; Boriboonsomsin et al., 2013]. This is due to the stronger aerodynamic drag force at higher speeds. In addition, some roadway sections with lighter traffic conditions may include steep grades, requiring more energy for the vehicle to climb the hills while producing more emissions in the process.

As part of an ECO-ITS research program sponsored by a variety of sources (e.g. automotive companies, California Department of Transportation, Federal Highway Administration, etc.), the research team has developed an eco-routing navigation technology [Barth et al., 2007;

Boriboonsomsin et al., 2010] for conventional fossil fuel-powered vehicles, both light-duty cars and heavy-duty trucks. However, For EVs, the benefits of the eco-routing navigation system are well beyond energy and emission savings. The technology can increase the possibility for an EV driver to complete his or her trip without running out of electricity, which helps relieve the "range anxiety" for an EV owner to some extent. This in turn could stimulate the purchase of EVs. As pointed out in a recent survey conducted by IBM Institute for Business Value for U.S. consumers [Gyimesi et al., 2010], "range anxiety"—the fear that an all-electric vehicle will not make it to desired destination before running out of power—is still thought to be one of the major obstacles to mass adoption of EVs. In addition, the application of eco-routing navigation systems to EVs can help provide guidance on the installation of public infrastructures needed for EVs, such as locations and types of recharging facilities.

Current commercialized EV navigation systems provide several functionalities. Most of them, such as the one on Nissan LEAF, are focused on locating a public charging-station in the vicinity, showing the current travel range of the vehicle, and displaying energy use by the engine and other components in real-time [1]. In terms of routing, the majority of these EV navigation systems still find a route corresponding to the shortest distance or shortest travel time between an origin and a destination. A few EV navigation systems in the market offer a similar eco-routing functionality, for example:
- Ford Focus Electric's onboard navigation system "provides an EcoRoute option based on characteristics of efficient EV driving." [2]
- Honda Fit EV's navigation system "can assist the user to find the best suitable route by using an interactive coaching system installed in the board." [3].

It is noted that, however, the algorithms underlying eco-route calculation in these OEM systems are proprietary and not publicly available. In addition, to the best of research teams' knowledge and technology assessment, none of them has a capability to incorporate real-time traffic information, road type, road grade, weight and/or weather information in the eco-route calculation. Therefore, as the subject of this project, to investigate the feasibility of obtaining the least energy consumed route for an EV under real-world conditions becomes very critical.

# Project Objectives

The goal of this project is to determine the feasibility of calculating a travel route for an electric vehicle (EV) that will require the least amount of energy for the trip, and thus extending the range of the EV.

To fulfil this goal, the research team has originally set up the following performance/cost objectives:

- Demonstrate that the calibrated simulation tool is able to calculate energy consumption of the test EV within an average error of ± 10%.
- Demonstrate that the algorithm is able to estimate energy consumption of the test EV in real-time within an average error of ± 10%.
- Perform field tests on an EV with prototype system versus without prototype system on both freeway and arterial roadways with various gradients for at least 50 trips.
- Demonstrate that the prototype system save the energy consumption of the test EV by an average of at least 10%.

The research team has modified the original statement of work due to the unexpected delays in obtaining a test electric vehicle. In addition, with the availability of CONSULT III plus Kit which is specifically designed to collect data and diagnostics systems for NISSAN LEAF, the research team can access a large amount of real-world high-resolution data (20 Hz) from the vehicle itself and monitor/record its energy consumption. Therefore, instead of calibrating EV simulation model, the research team developed a new model to estimate the energy consumption of the test vehicle under different speed levels, roadway type and road grade. And the team calibrated the Energy Operational Parameter Set (EOPS) of the model using the collected field data.

To summarize, the new project objectives were to

- Demonstrate that the data acquisition system is able to provide measurements within an average error of ± 10%.
    Demonstration of the accuracy of the data acquisition system is necessary to guarantee the validity of field data for EOPS calibration.
- Demonstrate that the algorithm is able to estimate energy consumption of the test EV in real-time within an average error of ± 10%.
    Demonstration of the algorithm estimation errors was desired in order to understand how well the prediction of link cost (link-based energy consumption rate) is.
- Perform field tests on an EV with prototype system versus without prototype system on both freeway and arterial roadways with various gradients for at least 50 trips.
    Conducting enough number of trips for the field testing was indispensable because: 1) the algorithm/model development requires enough data; 2) the data should cover a variety of scenarios to validate the prototype system.
- Demonstrate that the prototype system save the energy consumption of the test EV by an average of at least 10%.

Demonstration of the energy savings is indispensable because the project goal is to investigate the feasibility of obtaining the least energy consumption route for the test EV.

The following section will describe the methodology to achieve each project objective in details.

# Project Approach

To achieve the project goal, the research team identified the following tasks:
- Acquire the data acquisition systems and test electric vehicle (EV).
- Calibrate the energy operational parameter set (EOPS) for the test EV
- Develop real-time eco-routing algorithm for the test EV
- Implement the developed algorithm in prototype eco-routing navigation system for the test EV
- Finalize test plan
- Evaluate the prototype system

For each task, the team applied the associated approach as presented below.

**Acquisition of the Data Acquisition Systems and Test Electric Vehicle (EV)**

Rather than acquiring the simulation tool as scheduled originally, the research team has cooperated with NISSAN North America and obtained the specifically designed data acquisition system, CONSULT III plus Kit, which is capable of collecting data and diagnosing systems for NISSAN vehicles, such as LEAF. Therefore, instead of calibrating EV simulation model, the research team can develop a more reliable model to estimate the energy consumption of the test vehicle under a variety of situations.

More specifically, to collect the data of second-by-second vehicle activities (e.g., speed), energy consumption and geographic information (e.g., roadway grade), the research team used the following two data acquisition systems:
- CONSULT III plus kit (see Figure 1).
  As aforementioned, the CONSULT III plus kit can access high-resolution on-board diagnostics (OBD) data of a NISSAN LEAF, including vehicle speed, battery pack current (positive for battery pack charging while negative for discharging) and voltage, A/C power, accessory power, and etc. Therefore, the net propulsion power, $P^{prop}$, can be estimated as

$$P^{prop} = -\left(I^{bp} \times V^{bp}\right) - (P^{AC} + P^{acc})$$

  where $I^{bp}$ and $V^{bp}$ represent the instantaneous current (in ampere) and voltage (in volt) from/to the battery pack. $P^{AC}$ and $P^{acc}$ are the power consumed by air conditioner and other accessories (e.g., radio), respectively.
- GPS data logger (see Figure 2).
  It is well known that roadway grade is one of major contributing factors to a vehicle's energy consumption. The research team used the GPS data loggers to collect latitude and longitude information, and then synchronized with existing geographic information system (GIS) to acquire the roadway grade of each data point.

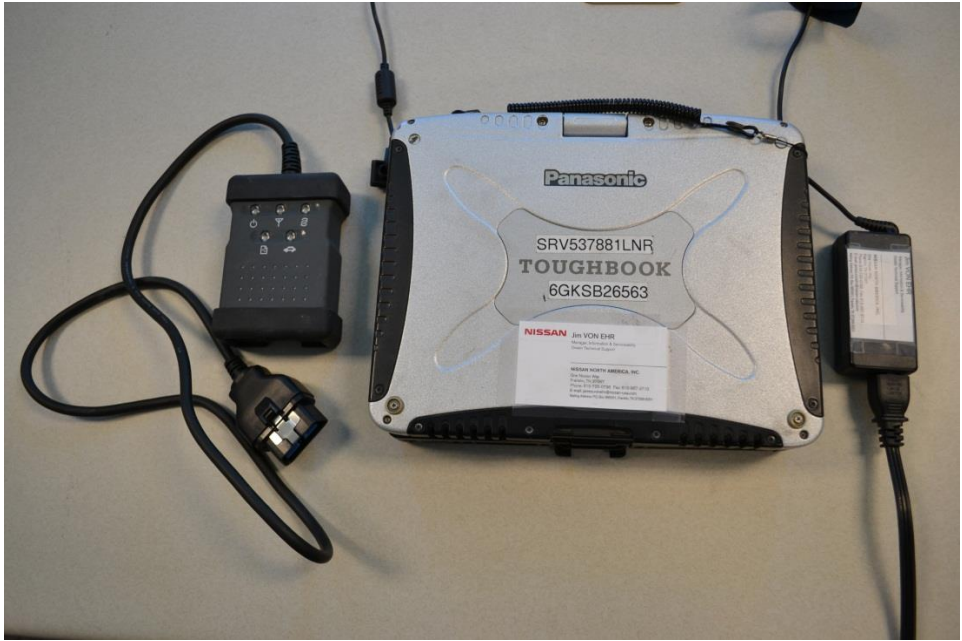

Figure 1. CONSULT III plus Kit to collect test NISSAN LEAF's energy consumption and other activity data

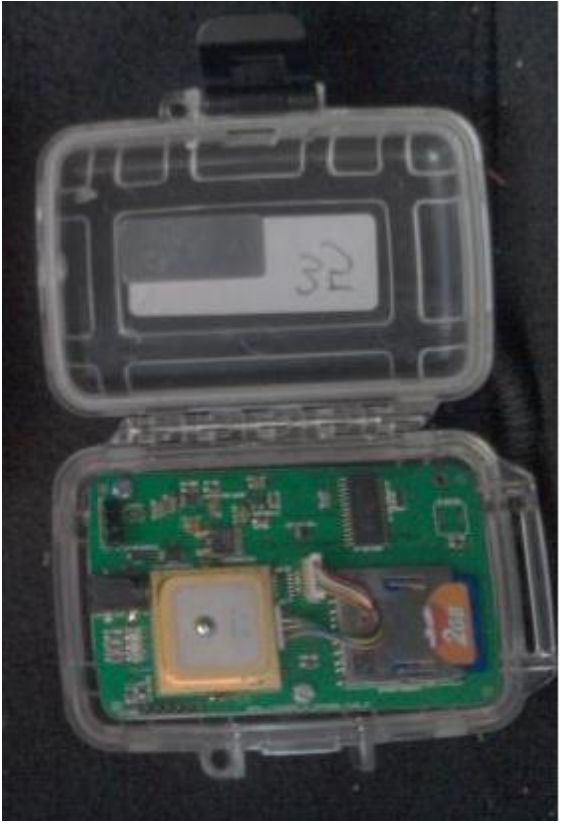

Figure 2. GPS data logger

On the other hand, with significant efforts in interactions with local NISSAN dealers and car rental companies, the research team succeeded in acquiring a 2013 NISSAN LEAF as the test electric vehicle (EV) and conducted field data collection from the end of March 2013 to July 2013. However, the difficulties in obtaining the test EV did delay the whole project schedule and the research team modified the original scope of work in order to catch up with the proposed project time frame. The modification will be further explained in the following sections.

*Collection of field data*

With the availability of data acquisition systems, the research team set up two types of experiments to collect the EV data in the field:
- Driving under the real-world conditions

    To cover a variety of traffic conditions, roadway types and grades, the research team carefully chose three pairs of origin-destination and routes (see Figure 3) to conduct real-world driving data collection for about four months. In total, the research team has successfully collected more than 100 trips (~1 hour each), which are non-trivial efforts, considering the schedules of driver(s) as well as the charging time and availability of the test EV.
- Driving under a controlled environment

    Besides the real-world driving experiment, the research team also collected data under a controlled environment for better understanding the energy consumption rates of the test EV at different cruise speeds with different operating modes (normal vs. eco). The cruise speed varies from 5 mph to 50 mph with 5 mph increment. Figure 4 presents the study site for control data collection, where there were very few interactions from other traffic during the data collection period.

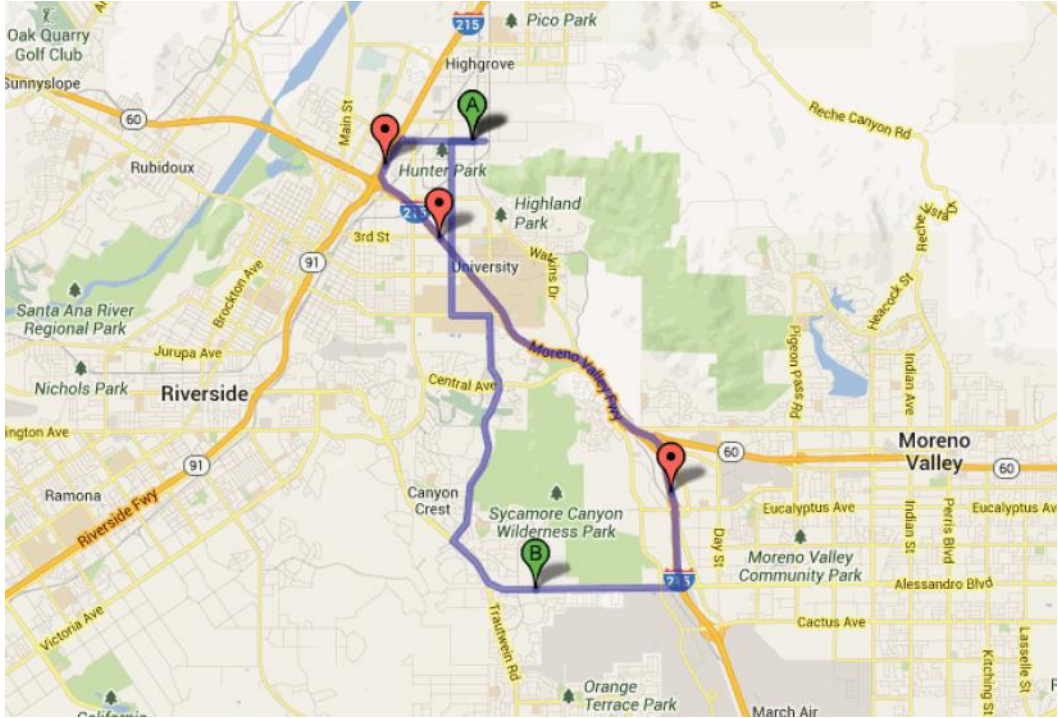

(a) Loop between CE-CERT and Alessandro Blvd., Riverside, CA

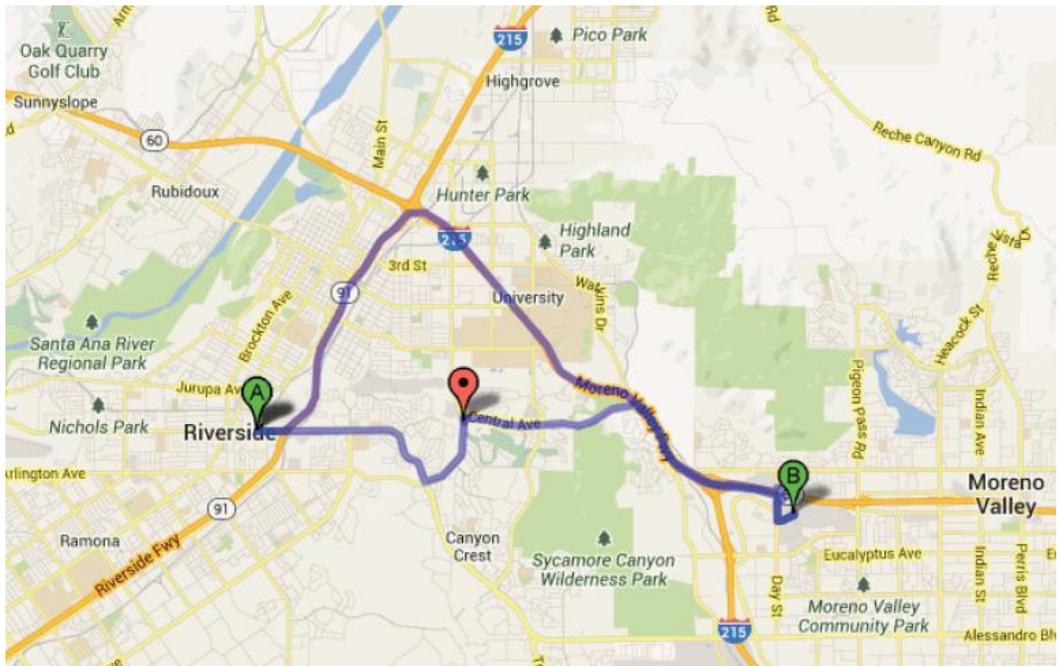

(b) Loop between Riverside Plaza and Towngate Circle, Moreno Valley, CA

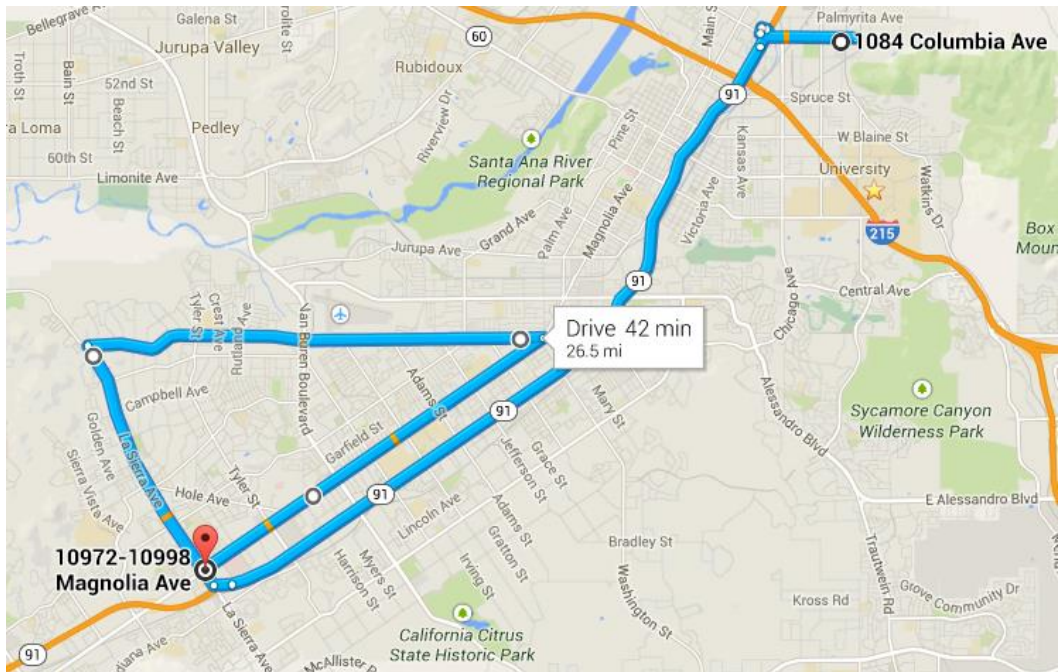
(c) Loop between CE-CERT and Magnolia Ave., Riverside, CA
Figure 3. Routes for real-world driving data collection

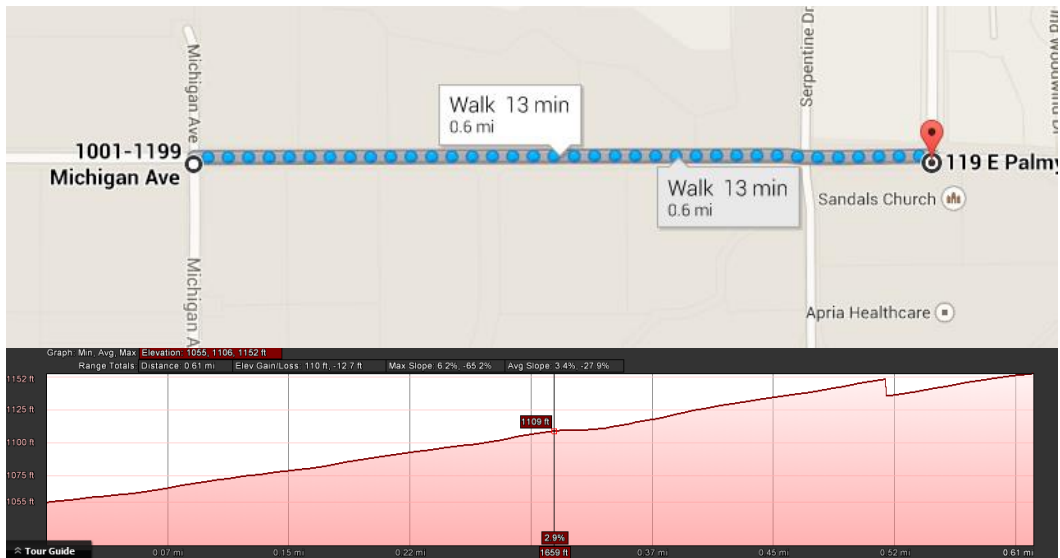
Figure 4. Routes for data collection under controlled environment

*Data fusion*

Before applying the field data to the development of EV eco-routing navigation algorithm, the research team has spent significant efforts in fusing data from both the CONSULT III plus kit and GPS data loggers. More specifically, there are two steps for data fusion in this project.
- Frequency realignment.
    The raw data files from GPS data loggers are not aligned with the ones from the CONSULT III plus kit in terms of updating rate. The research team has developed

routines in Python [4] to re-align the raw GPS data into 1 Hz, which is suitable for data synchronization as well as energy consumption estimation.
- Trip start time synchronization.

    It is noted that the GPS data logger is able to report the Coordinated Universal Time (UTC) as reference. However, CONSULT III plus kit only report the relative time stamp (i.e., each run always start from time "0") for each run. To fuse these two data sources, a common feature needs to be identified. In this project, the research team selected the vehicle speed information and applied cross-correlation technique [Billings, 2013] to synchronize these two data sources. The Project Outcome section will present an example for illustration.

Figure 5 summarizes the procedures to synchronize different data sources.

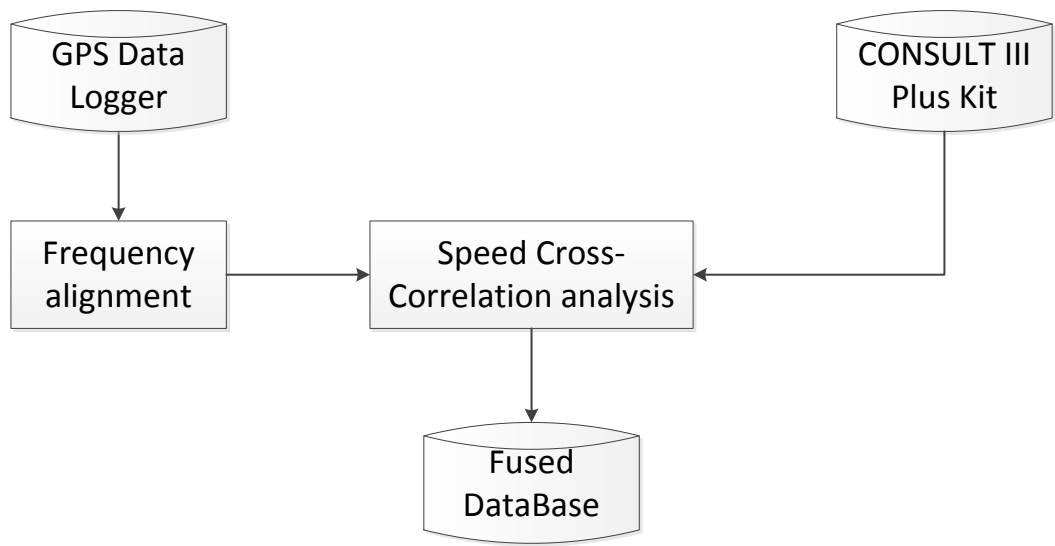

Figure 5. Flowchart for fusing data from the CONSULT III plus kit and GPS data loggers

**Calibration of the Energy Operational Parameter Set (EOPS) for the Test EV**

Due to the delays in acquiring a test EV and availability of field data, the research team modified the original Task 2 – "Calibrate the EV Simulation Tool for the Test EV" – to a new one "Calibrate the Energy Operational Parameter Set (EOPS) for the Test EV", where EOPS refers to the distance-based energy factors. For a specific electric vehicle (EV), several factors influence the energy consumption, including vehicle speed, roadway grade and roadway type. For example, compared to the downhill driving, an EV in general requires more energy when climbing a hill at the same speed level to overcome the gradient resistance. It is shown that the estimation of EOPS for conventional gasoline vehicle is quite complicated [Boriboonsomsin et al., 2012].

In this project, the research team conducted the following procedure (see Figure 6) to calibrate the EOPS of the test EV:

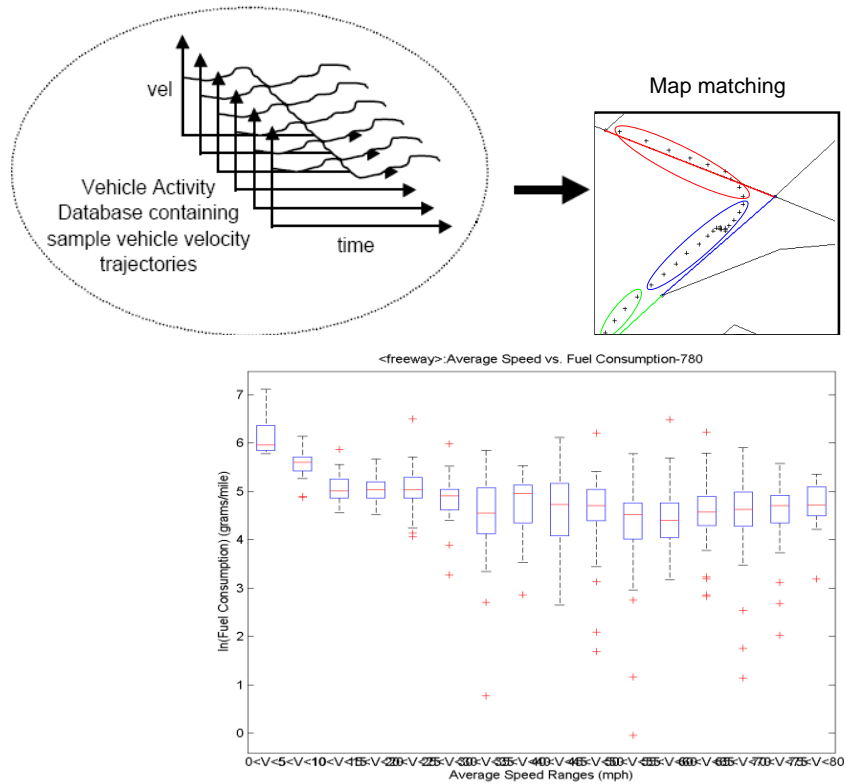

- Conduct map matching onto the fused dataset, where each data point is matched to the associated roadway link on the digital map that has the least orthogonal distance to the data point.
- Break down the fused data stream into short driving snippets based on roadway link.
- Group link-based snippets by roadway type and grade, allowing separate EOPS to be created for the different roadway types and grade, which will also improve the design of real-time eco-routing algorithm and facilitate the following tasks – "Develop Real-Time Eco-Routing Algorithm for the Test EV" and "Implement the Developed Algorithm in Prototype Eco-Routing Navigation System for the Test EV".
- Calculate the average speed and energy consumption rate (per mile) for each real-world driving snippet
- Apply parametric regression technique to model the statistics of distance-based energy consumption rate (i.e., median value) as a function of average speed of each snippet, taking into account the roadway type as well as roadway grade. It is noted that the box-and-whisker plots [McGill et al., 1978] are presented herein for the illustration purpose.

Using the new approach, the research team can better estimate the energy consumption for the customized EV. In addition, this new approach takes into account the fact that energy consumption data of the vehicle along with its speed and position will be monitored and archived continuously, which enables the on-line calibration of EOPS.

**Development and Implementation of Real-Time Eco-Routing Algorithm for the Test EV**

*Road network representation and real-time traffic information feeding*
Digital roadway networks consist of nodes (e.g., intersections, freeway on/off-ramps, point of curvature, etc.) and links (i.e., the road sections between nodes). Specific link and node attributes define how the network is connected together and what the general features are of the different links/nodes (e.g., position, length, number of lanes, capacity, speed limit, etc.). On the other hand, the traffic parameters (e.g., average traffic speed) on each link can be updated in real time (every 5 min), depending on the availability of traffic information.  Based on the attributes of each link (e.g., roadway type and grade), the associated EOPS (which is also a function of average traffic speed) is updated accordingly.

*Eco-routing engine*
Route calculation in the EV eco-routing navigation system is based on Dijkstra algorithm with the binary heap priority queue, which searches for a least cost path in a graph with non-negative edge path costs [Dijkstra, 1959]. Path construction can take into account users' route preference such as preferring highways or avoiding toll roads. In addition, it can take the number of passengers as an input for determining the vehicle's eligibility to use high-occupancy vehicle lanes. Note that, for each trip, the eco-routing navigation system can generate up to three route based on different minimization criteria, i.e., distance, travel time, and energy use.

*System architecture*
The overall architecture of the eco-routing navigation system is shown in Figure 7. First, the system receives a request from the driver through the user interface. Then, it triggers the EOPS  Updater to update EOPS for each link in the roadway network based on the latest set of fused traffic performance data from DynaNet and the calibrated EOPS for the test EV. Note that DynaNet can be set to periodically acquire and update traffic performance data (e.g., every 5 min) or on-demand (i.e., when there is a route request). Once the EOPS (in units of Watt-hour per unit distance) has been updated, it is multiplied by the link distance to yield the total energy estimates that are required so that the EV can traverse the link. These energy estimates are then stored in a data table in DynaNet along with other link cost factors (i.e., distance and estimated travel time). This data table, in conjunction with other data tables (such as the table that defines connectivity among links), is then used by the Routing Engine in calculating optimal routes according to the minimization criteria selected by the driver. Finally, the calculated routes are sent to the User Interface to be displayed to the driver. The system architecture has been implemented in two different fashions. In one system architecture has been implemented in two different fashions. In one system architecture, all the components are tightly integrated, whereas in the other system architecture, front-end applications only include the User Interface and are connected to other components housed in the back-end server, as depicted in Figure 7.

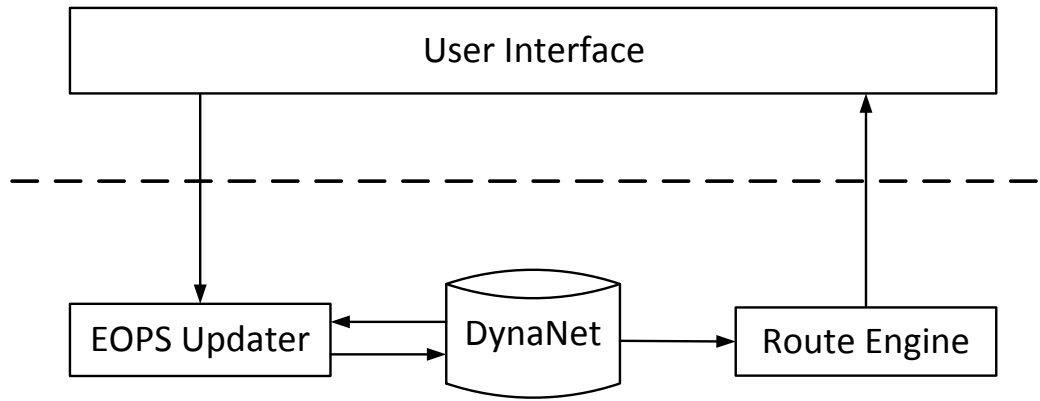

Figure 7. EV eco-routing system architecture

**Evaluation of Prototyping System**

After setting up the prototype system as aforementioned, the research team conducted a set of extensive simulation tests to evaluate the proposed algorithm. The testing period is from March 24th, 2014 to March 31th, 2014, 7 days covering Monday through Sunday. The testing origin-destination pairs are consistent with the ones for field data collection, i.e., Alessandro, Riverside Plaza and Magnolia. The real-world traffic information collected from the sensor (either fix-location or mobile) updated every 5 minutes and was archived in the server in CE-CERT at UC Riverside. In the simulation runs, the research team released three "virtual" vehicles at each origin which take 3 best routes (scenarios) in terms of shortest travel distance, shortest travel duration and least energy consumption, respectively. Therefore, there are totally [7 days * (24 hours/day) * (12 samples/hour) * 3 scenarios] = 6048 trips in the testing.

# Project Outcomes

Based on the predefined project objectives, the research team presents the outcome as follows.

**Objective 1: Demonstrate that the calibrated simulation tool is able to calculate energy consumption of the test EV within an average error of ± 10%.**

As aforementioned, the research team has modified the first objective to **demonstrate that the data acquisition system is able to provide measurements within an average error of ± 10%.**

The research team used the CONSULT III plus Kit, which is dedicated to the diagnostics of NISSAN vehicles, such as NISSAN LEAF, to acquired data. It has been confirmed with Tech Support that the measurements by using this system is quite accurate and are guaranteed within an average error of ± 10%.

The other source is GPS data logger. To test its measurement accuracy, the research team set the speed measurements from CONSULT III plus Kit as the benchmark and compared them with those from GPS data logger. However, as mentioned in the "Project Approach" section, the time stamps reported in CONSULT III plus Kit are not synchronized with the GPS time (or UTC time). The research team then used the speed data from both CONSULT III plus Kit and GPS data logger to sync both data source. Figure 8 presents an example of speed trajectory after synchronization.

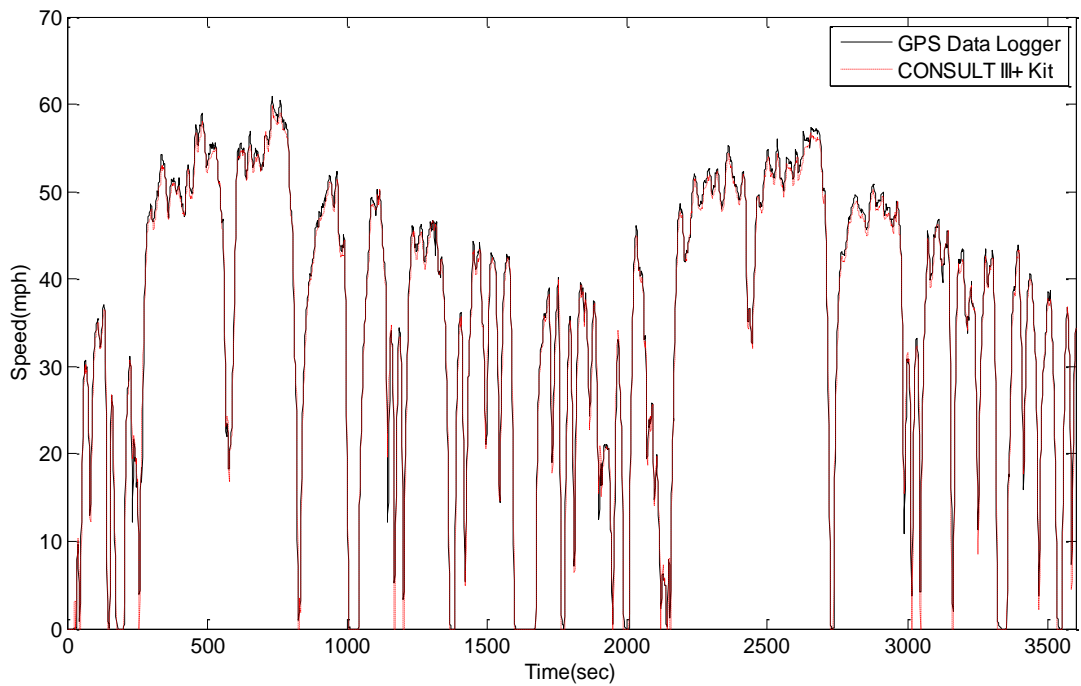

Figure 8. An example of speed trajectories from: GPS data logger vs. CONSULT III plus Kit

In addition, by comparing with the data from CONSULT III plus Kit, the research team quantified the percentage errors in terms of mean absolute percentage error (MAPE). However, there are a couple of samples with speed of zero. The symmetric mean absolute percentage error ([Armstrong, 1985] see below), or SMAPE, is used for evaluation.

$$SMAPE = \frac{\sum_{t=1}^{n}|F_t - A_t|}{\sum_{t=1}^{n}(A_t + F_t)}$$

where $n$ is the sample size; $A_t$ is the reference value (from CONSULTS III plus Kit in this case) and $F_t$ is the evaluated value (from GPS data logger in this case). Figure 9 presents the box-and-whisker plots of statistics of all SMAPE (all trips) for different study sites. It is obvious that the average (red bar) error metrics are below 1.2% across all study sites. In addition, the largest (upper end of the whisker) errors of the entire data sets are all slightly less than 1.7%. This demonstrates that the data acquisition system is able to provide measurements within an average error of ± 10%.

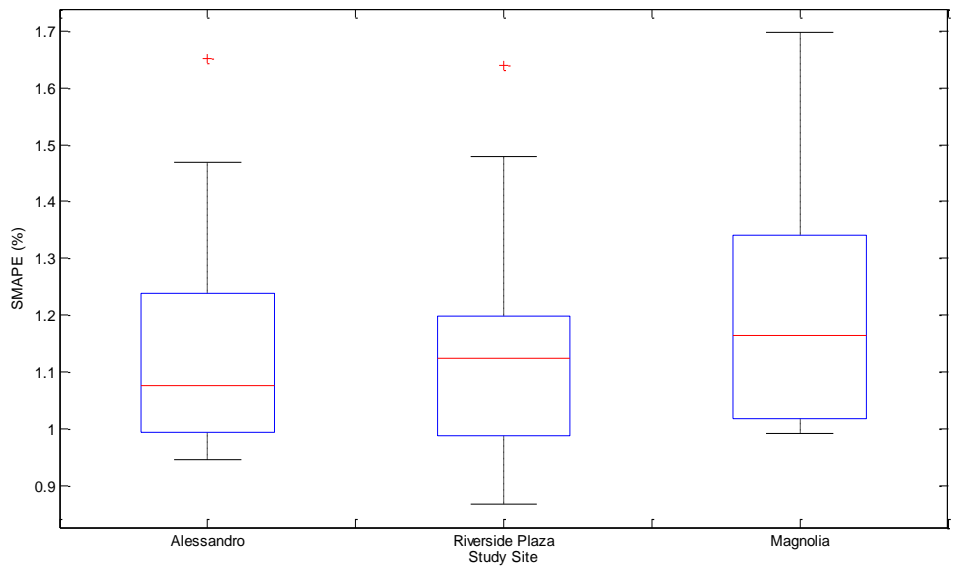

Figure 9. Boxplots of SMAPE (%) for the 3 study sites

**Objective 2: Demonstrate that the algorithm is able to estimate energy consumption of the test EV in real-time within an average error of ± 10%.**

By following the methodology described in "Project Approach" section, the research team obtained the boxplots of distance-based energy consumption vs. average speed bin (every 5 mph) at different roadway grade level. Figure 10 presents the EOPS at different speed level, including both freeway and arterial driving, when the roadway grade is 0. In addition, the research team conducted polynomial fits on the median values of distance-based energy consumption (vs. average speed) for different roadway grades, i.e.,

$$f_k = \sum_{i=0}^{4} \alpha_i \cdot v_k^i \qquad \text{(within certain roadway grade bin)}$$

where $i$ is the order of polynomial; $v_k$ is the (link-level) average vehicle speed (mph); $\alpha_i$'s are the associated coefficients in the regression model; and $f_k$ is the (link level) distance-based energy consumption (W-hr/mi). Figure 11 shows the results for roadway grade of 0. Unlike the typical EOPS "U-shaped" curve for gasoline light duty vehicle, the EOPS curve for the testing electric vehicle displays a "W-shape", where the drop within the low speed range (e.g., less than 30 mph) results from the regenerative braking. The research team further explored the negative kinetic energy (NKE) statistics [Watson et al., 1982] of each vehicle trajectory snippet in the entire data set to confirm the hypothesis. The NKE for each snippet is defined as

$$NKE = \frac{\sum_{i=1}^{N-1} \min(v_{i+1}^2 - v_i^2,\ 0)}{\sum_{i=1}^{N-1}(d_{i+1} - d_i)}$$

where $N$ is the length of snippet (in terms of time step); $v_i$ is the instantaneous speed (mph); and $d_i$ is the total travel distance up to the $i$-th time step. Figure 12 illustrates the NKE curve (using polynomial fit on median average speed) over the same data set. It can be seen that there is a significant drop between 0 mph and 30 mph, which represents frequent deceleration activities within this speed range. The energy in this spectrum can be well captured by the regenerative braking.

To understand how well the proposed algorithm can estimate the energy consumption of the testing EV in real-time, the research team chose the Riverside Plaza data set (an independent data set for validation), which is the largest data set covering both freeway and arterial (see Table 1), and calculated the error statistics (estimated route energy consumption vs. measured route energy consumption). Table 1 summarizes the error statistics, where the mean error in route-level energy consumption for the selected data set is -6.94%. It falls into the range of [-10%, 10%].

Table 1. Summary of error statistics of route-level energy consumption for Riverside Plaza data set

| # Trips | Error Statistics (%) | | | | | |
|---|---|---|---|---|---|---|
| | Min | 25-percentile | Mean | Median | 75-percentile | Max |
| 56 | -25.88 | -17.99 | -6.94 | -13.83 | 5.20 | 20.26 |

It is noted that other regression techniques, such as segmented regression (combination of logarithm form and polynomial form), can be applied to improve the model estimation capability at the very low speed value (e.g., around 0 mph).

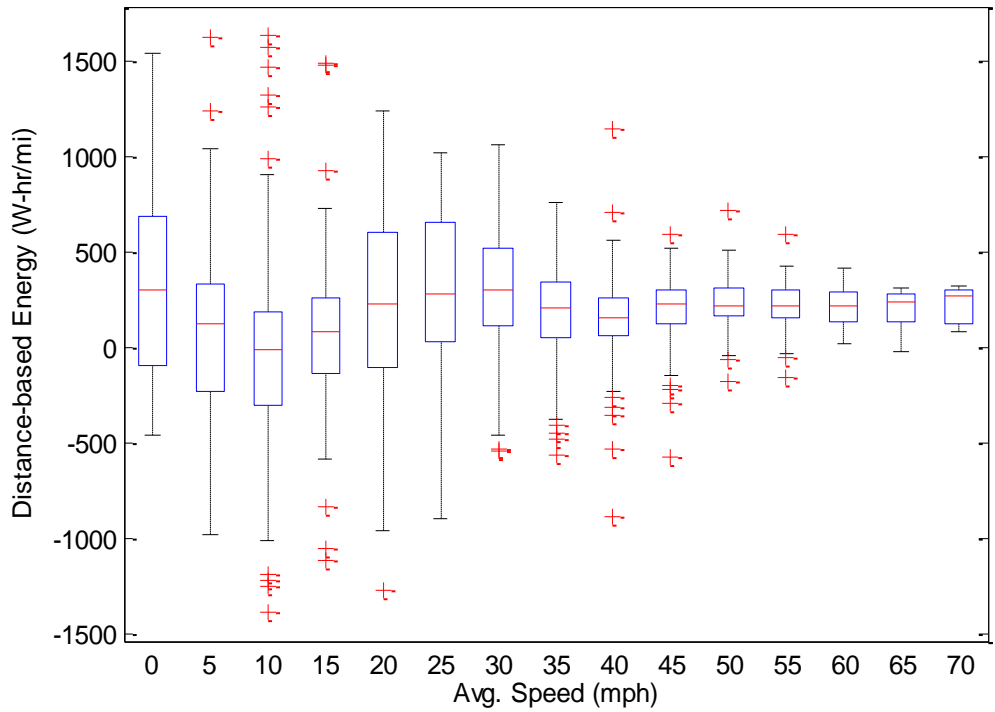

Figure 10. EOPS over different average speed for the entire data set (Normal mode)

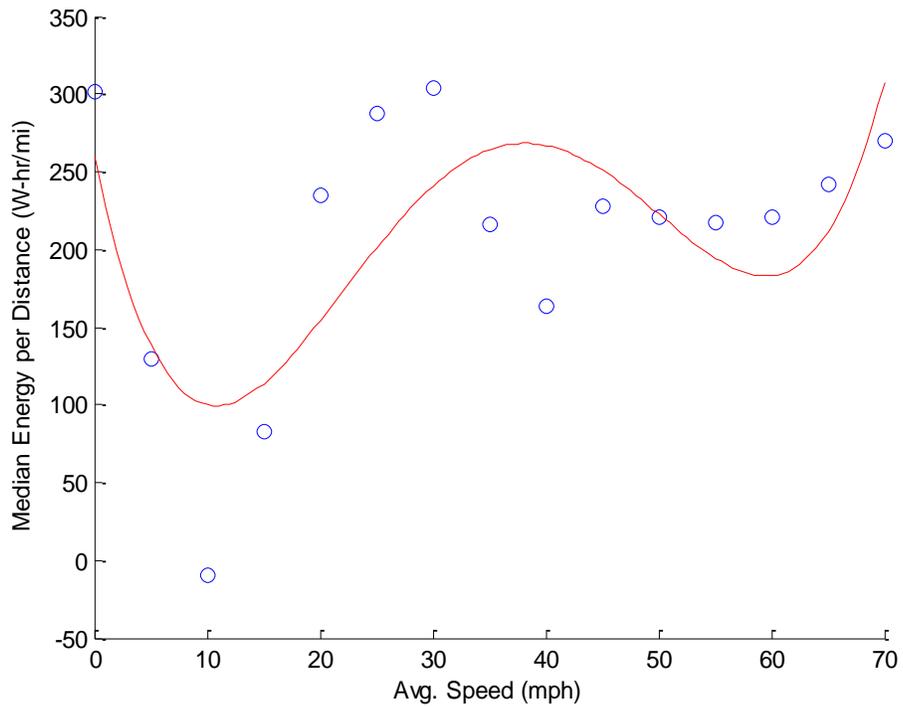

Figure 11. Polynomial fit of median EOPS over different average speed level

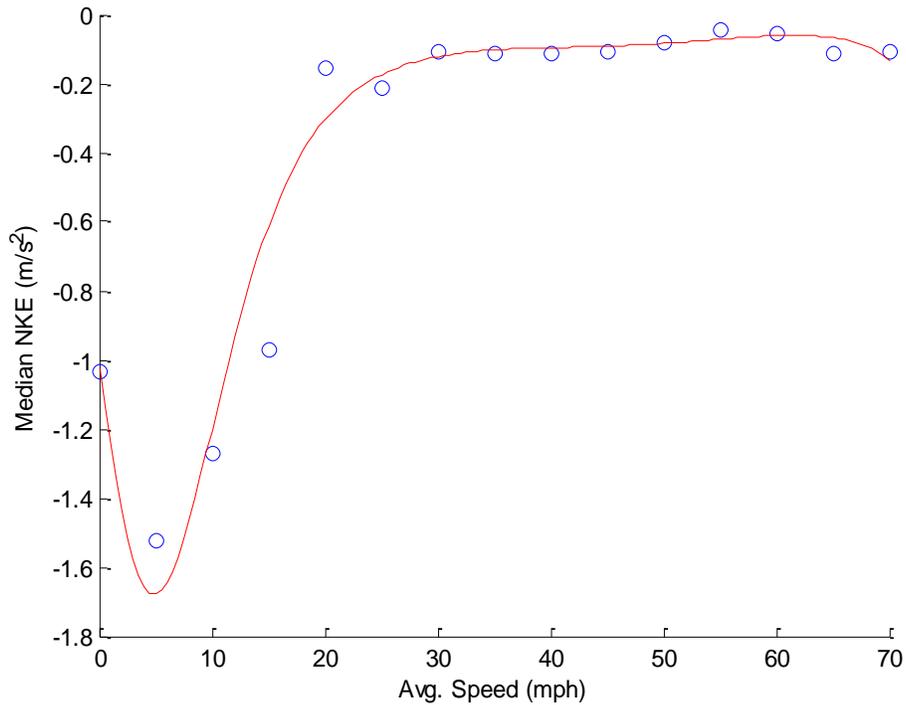

Figure 12. Polynomial fit of median NKE over different average speed level

**Objective 3: Perform field tests on an EV with prototype system versus without prototype system on both freeway and arterial roadways with various gradients for at least 50 trips.**

In order to collect enough data for model/algorithm development and validation, the research team performed extensive field tests between the selected pairs of origin and destination (see Figure 3). The numbers of trips for Alessandro, Riverside Plaza and Magnolia are 36, 40 and 56, respectively.

As described in the "Project Approach" section, the research team conducted two data collection methods:
1. Field data collection in actual traffic conditions. Table 2 and Table 3 summarize the key statistics of the data samples under different modes, i.e., normal mode and eco-mode. In total, the collected field data cover 3154 miles in distance, more than 100 hours (excluding Level 2 charging time) in time and over 70700 links (based on Geographic Information System).

Table 2. Summary of field data statistics under normal mode

| Study Site | Freeway | | | Arterial | | |
|---|---|---|---|---|---|---|
| | Miles (mi) | Seconds (s) | # of Sublinks | Miles (mi) | Seconds (s) | # of Sublinks |
| **Alessandro** | 0.00 | 0 | 0 | 543.36 | 63762 | 4786 |
| **Riverside** | 403.99 | 31944 | 16701 | 674.38 | 87595 | 11866 |
| Plaza | | | | | | |
| Magnolia | 312.25 | 26509 | 6452 | 299.06 | 49766 | 4496 |
| Total | 716.24 | 58453 | 23153 | 1516.80 | 201123 | 21148 |

Table 3. Summary of field data statistics under eco mode

| | Freeway | | | Arterial | | |
|---|---|---|---|---|---|---|
| **Study Site** | **Miles (mi)** | **Seconds (s)** | **# of Sublinks** | **Miles (mi)** | **Seconds (s)** | **# of Sublinks** |
| **Alessandro** | 0 | 0 | 0 | 0 | 0 | 0 |
| **Riverside Plaza** | 383.25 | 32549 | 16533 | 538.13 | 69866 | 9907 |
| **Magnolia** | 0 | 0 | 0 | 0 | 0 | 0 |
| **Total** | 383.25 | 32549 | 16533 | 538.13 | 69866 | 9907 |

2. Field data collection in control environment. Figure 13 presents the comparison results on distance based energy consumption (at different speed levels) between normal mode and eco mode. As can be observed from the figure, there are three major points:
   a. No matter what mode the driver selected, the roadway grade has significant impact on the energy consumption for the testing electric vehicle, due to the regenerative braking. The westbound trips (downhill) showed significantly lower distance-based energy consumption than the eastbound ones at any speed level.
   b. Under the driving at constant speed (which may not be the case in the real-world driving), the sweet points (where the least distance-based energy consumption occurs) for both Normal mode and Eco mode fall into the range between 10 mph and 20 mph. The most efficient speed is 16 mph and 13 mph for downhill and uphill, respectively.
   c. Compared to the Normal mode, the Eco mode shows much more benefits along the downhill terrain than along the uphill terrain, because the Eco mode is supposed to be more efficient in energy recovery from the regenerative braking.

It is noted that in the following analysis, the research team will focus on the Normal mode data set.

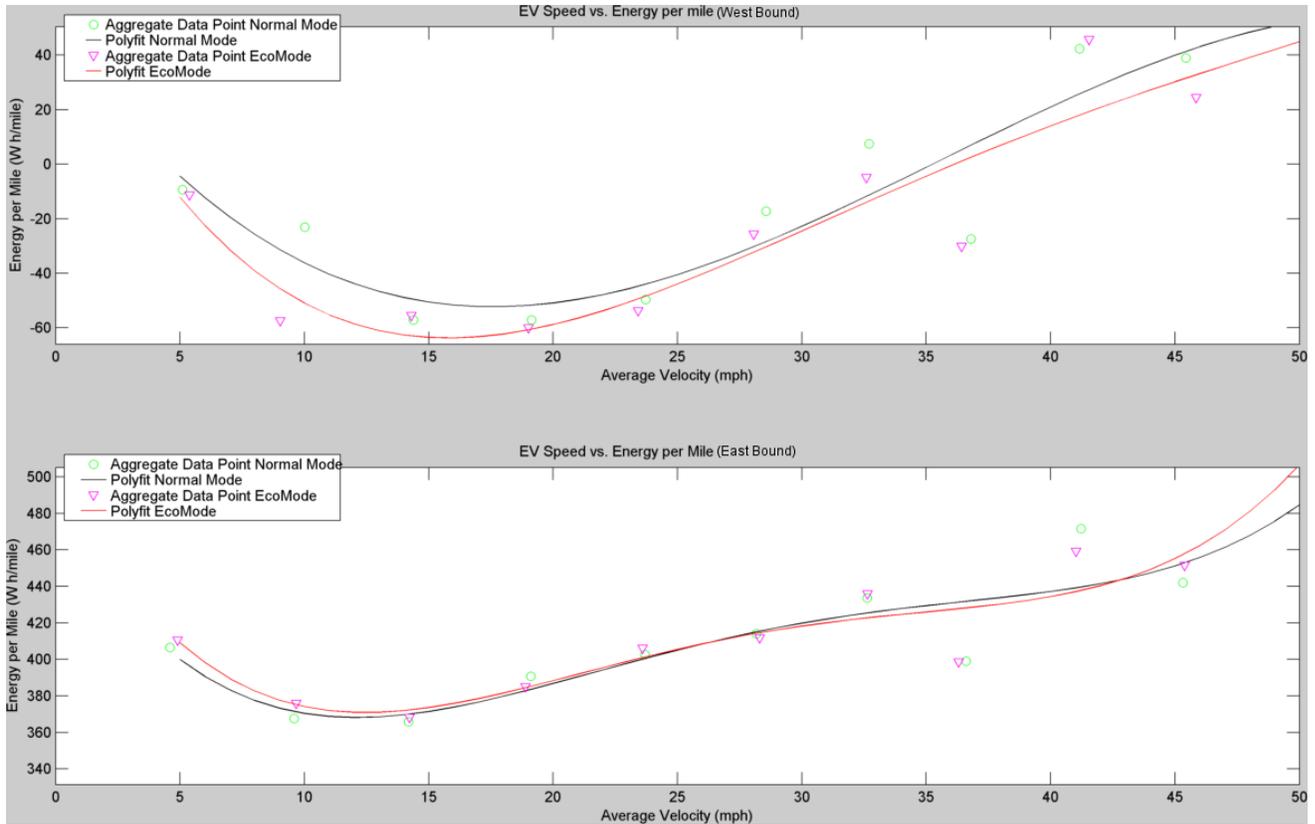
Figure 13. Comparison results from the control experiment

**Objective 4: Demonstrate that the prototype system save the energy consumption of the test EV by an average of at least 10%.**

As illustrated in the "Project Approach", the research team conducted a set of extensive simulation tests. Figure 14 provides an example of 3 best routes starting at the same time. As can be seen from the figure, shortest travel distance route and least travel duration route overlap a lot, but the least energy consumption route is quite different from the other two.

Table 4 summarizes the results from the simulation test. It is shown that compared with the least energy consumption route, the average relative increments in energy for the shortest distance routes range from 5% to 25%, depending on the study sites and travel direction. Such range for the least duration routes becomes wider (between 25% and 51%). This demonstrates that the prototype system can save the energy consumption of the test EV by more than 10% on average. In addition, the least energy consumption routes penalize on both travel distance (between 11% and 22%) as well as travel time (more than 97%), because the sweet spot is around 12 mph (see Figure 11).

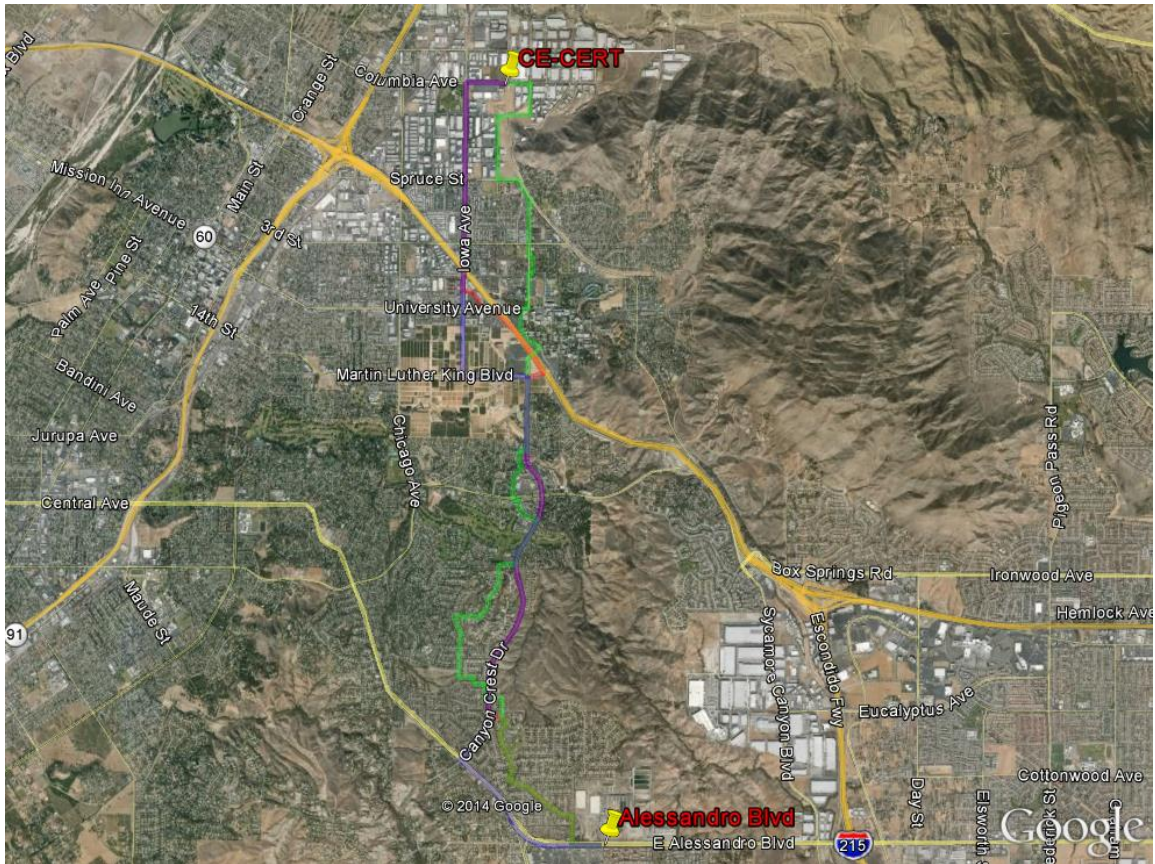

Figure 14. An example of 3 routes between CE-CERT and Alessandro Blvd, Riverside, starting at the same time: red – shortest travel distance; blue – least travel time; and green – least energy consumption

Table 4. Summary of results from the simulation test for 3 different route choice scenarios

| Study Sites | Direction | Mean Relative Increment (%) for Different Scenarios | | | | | |
|---|---|---|---|---|---|---|---|
| | | Shortest Distance | | Least Duration | | Least Energy | |
| | | Time Increment | Energy Increment | Distance Increment | Energy Increment | Distance Increment | Time Increment |
| 1 | 0 | 22.7 | 15.1 | 16.9 | 44.6 | 10.8 | 97.4 |
| | 1 | 57.3 | 5.0 | 19.0 | 50.6 | 11.1 | 109.9 |
| 2 | 0 | 7.6 | 19.0 | 16.7 | 26.2 | 22.1 | 185.5 |
| | 1 | 10.6 | 25.0 | 9.1 | 27.3 | 15.0 | 146.5 |
| 3 | 0 | 70.4 | 14.3 | 3.3 | 24.5 | 16.2 | 186.4 |
| | 1 | 95.7 | 9.3 | 3.9 | 30.0 | 13.1 | 175.1 |

# Conclusions

Based on the factual findings from the previous sections, the research team can safely draw the following conclusions:
1. As is shown in the Project Outcome section, the data acquisition systems implemented in this project, including the CONSULT III plus Kit and GPS data logger, are quite reliable and can satisfy the requirement. The difference in speed measurement between the GPS data logger and dedicated diagnostics system (i.e., CONSULTS III plus Kit) across all study sites are less than 1.7% on average. This lay the foundation for further development of the algorithm;
2. The proposed polynomial regression model on the Energy Operational Parameter Set (EOPS), despite simplicity, is able to estimate energy consumption of the test EV in real-time within an average error of ± 10%. For example, the mean error (%) in estimation of trip-level energy consumption is only 0.6%, when the research team applied to the Riverside Plaza study site;
3. The results have indicated that the number of trips identified in the scope of work is enough and the data set is reliable for algorithm development and validation. In addition, field tests showed that: a) the roadway grade plays a significant role on energy consumption for the electric vehicle, due to the introduction of regenerative braking; b) the eco mode outperforms the normal mode when the route experiences downhill terrain, because the eco mode is more efficient in energy recovery;
4. Extensive evaluation tests on the selected study sites show that the prototype system provides significant savings in energy consumption, compared to conventional navigation systems, which has indicated significant potential for commercialization. However, there may be certain penalties in travel time and travel distance.

# Recommendations

Upon the completion of this project, the research team has a couple of recommendations:
1. The results are site specific. As can be shown in the previous section, the system performance may vary significantly with study sites (i.e., the origin-destination pairs) and even travel directions. Therefore, a larger scale testing under a more generalized framework should be conducted. As suggested in [Boriboonsomsin et al, 2013], the research team can perform a more extensive evaluation by gridding a larger network and defining the origin-destination pair based on grids (see Figure 15).

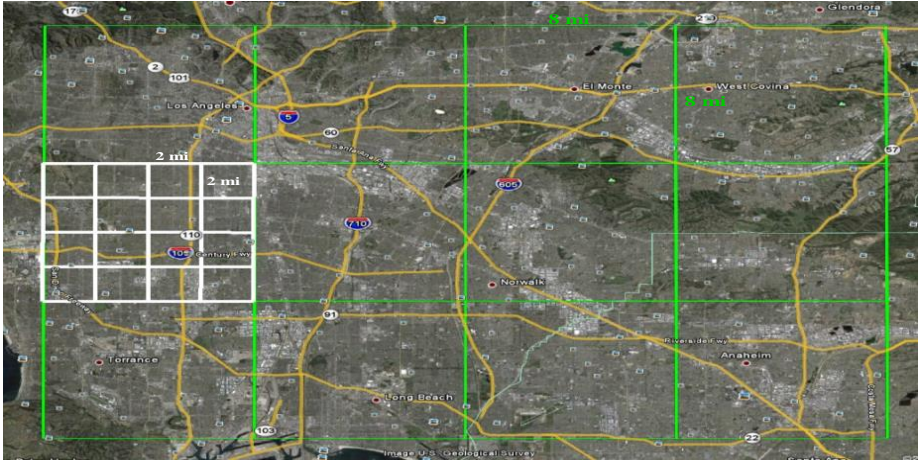

Figure 15. An alternative way to evaluate the system performance using grids for defining origins and destinations of artificial trips

2. If more resources are available, additional data should be collected to determine the impacts of different influential factors, such as vehicle model (rather than NISSAN LEAF) and behaviors of drivers (different drivers).
3. To get better understanding the characteristics of differences route choice, further analysis should be done by incorporating other information, such as time of the day or day of the week. Drivers' values on the travel time and energy may vary, depending on whether the trip is performed on a weekday or a weekend.
4. As shown in the previous section, the least energy consumption routes for the test EV usually have much higher penalty on the travel time (but not necessarily on the travel distance). For commercialization, how to select a more rational navigation objective function is worthy of further investigating to achieve the balance between energy consumption and travel time or travel distance and fulfill the customers' needs.

# Public Benefits to California

It is widely expected that there will be a significant growth in the EV market in the coming years. For example, by the end of Year 2013, NISSAN has already sold 42,000 units of NISSAN LEAF in U.S. [5]. And NISSAN projected to sell another 30,000 units in Year 2014 (U.S. only) [6]. Assume that 10% of them are operated in California. The current Nissan LEAF is estimated to consume 0.24 kWh per mile. Given the average annual mileage of 12,000, one Nissan LEAF will consume 2,880 kWh annually.

The outcome of this project shows that the energy savings can be as much as 51% by using the proposed navigation algorithm rather than the conventional least duration strategy. Therefore, as a result of using the eco-routing navigation technology, the 7,200 Nissan EVs (projected to the end of Year 2014) in California would translate to approximately 20,736,000 kWh per year. At $0.2 per kWh [7], the proposed navigation system would result in an annual saving of $4.15 million in electricity costs in California, just by Nissan LEAF EVs alone. If EVs from other manufacturers also implement the eco-routing navigation technology, the energy and cost savings will be even greater.

# References

Ahn, K. and H. Rakha (2008). "The Effects of Route Choice Decisions on Vehicle Energy Consumption and Emissions." Transportation Research Part D. pp. 151 – 167.

Armstrong, J. S. (1985). "Long-range Forecasting: From Crystal Ball to Computer." 2nd. ed. Wiley.

Barth, M. and K. Boriboonsomsin (2008). "Real-World CO2 Impacts of Traffic Congestion", Transportation Research Record No. 2058, pp 163-171, Transportation Research Board, National Academy of Science.

Barth, M., K. Boriboonsomsin and A. Vu (2007). "Environmental-Friendly Navigation." Proceedings of the 10th International IEEE Conference on Intelligent Transportation System (CD-ROM) Seattle, WA, September 30 – October 3.

Boriboonsomsin, K., M. Barth, W. Zhu and A. Vu (2010). "ECO-Routing Navigation System based on Multi-Source Historical and Real-Time Traffic Information" IEEE ITSC 2010 Workshop on Emergent Cooperative Technologies in Intelligent Transportation Systems.

Boriboonsomsin, K., M. Barth, W. Zhu and A. Vu (2012). "Eco-Routing Navigation System Based on Historical and Real-Time Traffic Information." IEEE Transactions on Intelligent Transportation Systems, Vol. 13, No. 4, pp. 1694 – 1704

Boriboonsomsin, K., J. Dean and M. Barth (2013). "An Examination of the Attributes and Value of Eco-Friendly Route Choices." Accepted by Transportation Research Record.

Billings, S.A. (2013) "Nonlinear System Identification: NARMAX Methods in the Time, Frequency, and Spatio-Temporal Domains". Wiley

Dijkstra, E. W. (1959). "A note on two problems in connexion with graphs". Numerische Mathematik 1: 269–271

Eberle, U. and R. von Helmolt (2010). "Sustainable transportation based on electric vehicle concepts: a brief overview" Energy Environmental Science, March 2010, pp. 689-699.

European Commission (2011). "Transport: Electric Vehicles – European Commission", Accessed on January 24, 2011, http://ec.europa.eu/transport/urban/vehicles/road/electric_en.htm

Gyimesi, K., S. Schumacher, J. Diehlmann, and S. Tellouck-Canel. (2010). "Advancing mobility: The new frontier of smarter transportation." IBM Institute for Business Value, ftp://public.dhe.ibm.com/common/ssi/ecm/en/gbe03375usen/GBE03375USEN.PDF.

McGill, R., J. W. Tukey and W. A. Larsen (1978). "Variations of Box Plots". The American Statistician 32 (1): 12–16

Watson, H. C. et al. (1982). "Development of the Melbourne Peak Cycle." Paper #82148, SAE of Australia

Wu, G., K. Boriboonsomsin, W.-B. Zhang, M. Li, and M. Barth. (2010). "Energy and emission benefit comparison between stationary and in-vehicle advanced driving alert systems." Transportation Research Record: Journal of the Transportation Research Board. No. 2189, 2010, pp. 98 – 106

[1] http://www.allcarselectric.com/news/1047586_video-2011-nissan-leaf-ev-navigation-system-demo

[2] http://electric-vehicles-cars-bikes.blogspot.com/2011/04/ford-ev-buzz-heats-up-as-chief-engineer.html

[3] http://www.electriccarsforsale.biz/2012-honda-fit-ev-new-electric-car-innovation.html

[4] https://www.python.org/download/releases/3.4.0/

[5] http://en.wikipedia.org/wiki/Nissan_Leaf

[6] http://insideevs.com/monthly-plug-in-sales-scorecard/

[7] http://www.bls.gov/ro9/cpilosa_energy.htm

# Development Status Questionnaire

| | |
|---|---|
| California Energy Commission<br>Energy Innovations Small Grant (EISG) Program<br>**PROJECT DEVELOPMENT STATUS** | **Questionnaire** |

Answer each question below and provide brief comments where appropriate to clarify status. If you are filling out this form in MS Word the comment block will expand to accommodate inserted text.

| Please Identify yourself, and your project: **PI Name** ___Guoyuan Wu___ **Grant** # _11-01T_____ ||
|---|---|
| **Overall Status** ||
| **Questions** | **Comments:** |
| 1) Do you consider that this research project proved the feasibility of your concept? | *Yes. As shown in the experiment, there exists eco-friendly route which is different from the shortest duration route for an EV. In addition, the resultant energy saving is up to 51%* |
| 2) Do you intend to continue this development effort towards commercialization? | *Yes. It is very promising and the research topic still needs further development effort before commercialization.* |
| **Engineering/Technical** ||
| 3) What are the key remaining technical or engineering obstacles that prevent product demonstration? | *To develop/improve the short-term traffic state prediction for on-line (real-time) implementation.* |
| 4) Have you defined a development path from where you are to product demonstration? | *Yes.* |
| 5) How many years are required to complete product development and demonstration? | *1 – 2 year(s)* |
| 6) How much money is required to complete engineering development and demonstration? | *About $200,000* |
| 7) Do you have an engineering requirements specification for your potential product? | *No, but we expect to have it completed in 2 years* |
| **Marketing** ||
| 8) What market does your concept serve? | *Both residential and commercial* |
| 9) What is the market need? | *To boost the mass adoption of EVs by relieving the "range anxiety"* |
| 10) Have you surveyed potential customers for interest in your product? | *No* |
| 11) Have you performed a market analysis that takes external factors into consideration? | *No* |
| 12) Have you identified any regulatory, institutional or legal barriers to product acceptance? | *No* |
| 13) What is the size of the potential market in California for your proposed technology? | *Take NISSAN as an example, the estimated share in California is expected to be 24,200 by Year 2015* |
| 14) Have you clearly identified the technology that can be patented? | *Yes* |
| 15) Have you performed a patent search? | *Yes. Self-search. But I did not find a similar yet.* |
| 16) Have you applied for patents? | *No* |
| 17) Have you secured any patents? | *No* |
| 18) Have you published any paper or publicly disclosed your concept in any way that would limit your ability to seek patent protection? | *No* |
| **Commercialization Path** | |
| 19) Can your organization commercialize your product without partnering with another organization? | *No. A auto-maker or navigation system maker would be ideal as a partner.* |
| 20) Has an industrial or commercial company expressed interest in helping you take your technology to the market? | *So far no.* |
| 21) Have you developed a commercialization plan? | *Not yet.* |
| 22) What are the commercialization risks? | *Popularization of EVs (which may be constrained due to the battery technology and availability of charging facilities)* |
| **Financial Plan** | |
| 23) If you plan to continue development of your concept, do you have a plan for the required funding? | *No* |
| 24) Have you identified funding requirements for each of the development and commercialization phases? | *No* |
| 25) Have you received any follow-on funding or commitments to fund the follow-on work to this grant? | *No. Potential sources may include auto-makers, navigation system manufacturers or government agencies* |
| 26) What are the go/no-go milestones in your commercialization plan? | *Not available* |
| 27) How would you assess the financial risk of bringing this product/service to the market? | *Conduct survey on the potential market* |
| 28) Have you developed a comprehensive business plan that incorporates the information requested in this questionnaire? | *No* |
| **Public Benefits** | |
| 29) What sectors will receive the greatest benefits as a result of your concept? | *Residential and environment* |
| 30) Identify the relevant savings to California in terms of kWh, cost, reliability, safety, environment etc. | If market share of NISSAN LEAF can hit 24,200 vehicles in California by Year 2015, implementation of the proposed system can annually save approximately 35,500,000 kWh in electricity, or $7.1 million (assuming $0.2 per kWh) in electricity costs. |
| 31) Does the proposed technology reduce emissions from power generation? | *The proposed technology may have secondary effects in reducing emission from power generation, because it may save the total energy in transportation.* |
| 32) Are there any potential negative effects from the application of this technology with regard to public safety, environment etc.? | *A system-wide navigation strategy needs to be further developed to avoid potential bottlenecks resulting from an ego-navigation strategy.* |
| **Competitive Analysis** | |
| 33) What are the comparative advantages of your product (compared to your competition) and how relevant are they to your customers? | *1. Specific to potential EVs buyers and stimulate the mass adoption of EVs;*<br>*2. Incorporated GIS and real-time traffic information when estimating the energy consumption;*<br>*3. It provides a generalized framework to Eco-Routing but not restricted to any specific make or model of EVs.* |
| 34) What are the comparative disadvantages of your product (compared to your competition) and how relevant are they to your customers? | *1. Functions of the proposed prototype system need to be much further polished before commercialization.*<br>*2. More detailed development is required to customize the product for specific EV model and customers' needs. For example, the product can provide multiple route options based on travelers' preference and recommendation by self-learning human behavior.*<br>*3. The route calculation time may be improved by applying more advanced algorithms to enhance users' experience.* |
| **Development Assistance** ||
| The EISG Program may in the future provide follow-on services to selected Awardees that would assist them in obtaining follow-on funding from the full range of funding sources (i.e. Partners, PIER, NSF, SBIR, DOE etc.). The types of services offered could include: (1) intellectual property assessment; (2) market assessment; (3) business plan development etc. ||
| 35) If selected, would you be interested in receiving development assistance? | *Yes.* market assessment and business plan development |